\documentstyle[prl,aps,epsfig,multicol]{revtex}
\begin{document}
\draft

\newcommand{\snn}{\sqrt{s_{NN}}}
\newcommand{\pp}{pp}
\newcommand{\pbarp}{\overline{p}p}
\newcommand{\zvtx}{z_{vtx}}
\newcommand{\spec}{SPEC}
\newcommand{\specp}{{\spec}P}
\newcommand{\specn}{{\spec}N}
\newcommand{\vtx}{VTX}
\newcommand{\vtxt}{{\vtx}T}
\newcommand{\vtxb}{{\vtx}B}

\newcommand{\np}{N_{part}}
\newcommand{\avenp}{\langle N_{part} \rangle}
\newcommand{\npB}{N_{part}^B}
\newcommand{\nc}{N_{coll}}
\newcommand{\half}{\frac{1}{2}}
\newcommand{\halfnp}{(\half\avenp)}

\newcommand{\nch}{N_{ch}}
\newcommand{\etazero}{\eta = 0}
\newcommand{\etaone}{|\eta| < 1}
\newcommand{\dndeta}{d\nch/d\eta}
\newcommand{\dndetazero}{\dndeta|_{\etazero}}
\newcommand{\dndetaone}{\dndeta|_{\etaone}}
\newcommand{\dndetanp}{\dndeta / \halfnp}
\newcommand{\dndetaonp}{\dndeta / \np}
\newcommand{\dndetazeronp}{\dndetazero / \halfnp}
\newcommand{\dndetaonenp}{\dndetaone / \halfnp}

\title{Centrality Dependence of Charged Particle Multiplicity
at Mid-Rapidity in Au+Au Collisions at $\snn =$ 130 GeV }
\author{ B.B.Back$^1$, M.D.Baker$^2$, 
D.S.Barton$^2$, R.R.Betts$^{6}$, R.Bindel$^7$,  
A.Budzanowski$^3$, W.Busza$^{4}$, A.Carroll$^2$,
M.P.Decowski$^4$, 
E.Garcia$^7$, N.George$^1$, K.Gulbrandsen$^4$, 
S.Gushue$^2$, C.Halliwell$^6$,
G.A.Heintzelman$^2$, C.Henderson$^4$, R.Ho\l y\'{n}ski$^3$, D.J.Hofman$^6$,
B.Holzman$^6$, 
E.Johnson$^8$, J.L.Kane$^4$, J.Katzy$^4$, N. Khan$^8$, W.Kucewicz$^{6}$, P.Kulinich$^4$,
W.T.Lin$^5$, S.Manly$^8$,  D.McLeod$^6$, J.Micha\l owski$^3$,
A.C.Mignerey$^7$, J.M\"ulmenst\"adt$^{4}$,R.Nouicer$^6$, 
A.Olszewski$^{2,3}$, R.Pak$^2$, I.C.Park$^8$, 
H.Pernegger$^4$, C.Reed$^4$, L.P.Remsberg$^2$, 
M.Reuter$^6$, C.Roland$^4$, G.Roland$^4$, L.Rosenberg$^4$, 
P.Sarin$^4$, P.Sawicki$^3$, 
W.Skulski$^8$, 
S.G.Steadman$^4$, 
G.S.F.Stephans$^4$, P.Steinberg$^2$, M.Stodulski$^3$, A.Sukhanov$^2$, 
J.-L.Tang$^5$, R.Teng$^8$, A.Trzupek$^3$, 
C.Vale$^4$, G.J.van Nieuwenhuizen$^4$, 
R.Verdier$^4$, B.Wadsworth$^{4}$, F.L.H.Wolfs$^8$, B.Wosiek$^3$, 
K.Wo\'{z}niak$^3$, 
A.H.Wuosmaa$^1$, B.Wys\l ouch$^4$\\
(PHOBOS Collaboration)\\
$^1$ Physics Division, Argonne National Laboratory, Argonne, IL 60439-4843\\
$^2$ Chemistry and Collider-Accelerator Departments, Brookhaven National Laboratory, Upton, NY 11973-5000\\
$^3$ Institute of Nuclear Physics, Krak\'{o}w, Poland\\
$^4$ Laboratory for Nuclear Science, Massachusetts Institute of Technology, Cambridge, MA 02139-4307\\
$^5$ Department of Physics, National Central University, Chung-Li, Taiwan\\
$^6$ Department of Physics, University of Illinois at Chicago, Chicago, IL 60607-7059\\
$^7$ Department of Chemistry and Biochemistry, University of Maryland, College Park, MD 20742\\
$^8$ Department of Physics and Astronomy, University of Rochester, Rochester, NY 14627
}

\date{\today}
\maketitle

\begin{abstract}
We present a measurement of the pseudorapidity density of 
primary charged particles near mid-rapidity in Au+Au collisions 
at $\snn =$ 130 GeV as a function of the number of participating
nucleons.  These results are compared to models in an attempt to
discriminate between
competing scenarios of particle production in heavy ion collisions.
\end{abstract}

\begin{multicols}{2}
\narrowtext

Collisions of gold nuclei at the Relativistic Heavy Ion 
Collider (RHIC) provide a unique opportunity to study particle production
in nuclear collisions at the highest available energies.  
In a previous publication \cite{dndeta}, the PHOBOS collaboration 
presented results on the energy dependence of 
the pseudorapidity density of charged particles, $\dndeta$, produced
near midrapidity for central Au+Au collisions.
It showed that this rises much faster with energy than in $\pbarp$ collisions
at similar energies\cite{pp}.
This has been explained by the increasing role of 
hard and semi-hard processes, which are understood using perturbative
QCD.

A way to control the ratio of hard to soft production at
a fixed beam energy is to vary the impact parameter, 
or centrality, of the nuclear collisions.
Soft processes, which produce the bulk of charged particles in
$\pp$ and $pA$ collisions, scale with the number of participating nucleons
($\np$) in the collision\cite{elias,wounded}.
Hard processes occur in the interactions between individual
partons in the colliding nucleons and are expected to scale
with the number of nucleon-nucleon collisions, $N_{coll}$.  
This leads to an expected scaling of $\dndetazero$ as 
$A \times \np +  B \times \nc$.

Data from the WA98 experiment at CERN \cite{wa98} already indicate possible
deviations from simple $\np$ scaling even at SPS energies ($\snn = 17.2$ GeV).
A stronger-than-linear dependence 
on $\np$ is observed, 
well described by a power-law, 
$\dndetazero \propto \np^\alpha $, with $\alpha = 1.07 \pm 0.04$.

Theoretical models of particle production in RHIC collisions generally
fall into two classes.
The first incorporates the expected scaling mentioned
above, using a Glauber model calculation \cite{glauber} 
to determine the relationship between
$\np$ and $\nc$ as a function of impact parameter.
The HIJING model \cite{hijing-cent} as well as calculations by
Kharzeev and Nardi (KN) \cite{kn} follow this approach.
HIJING also incorporates jet quenching and nuclear shadowing which
modifies the scaling, leading to a 
linear rise of the normalized multiplcity $\dndetaonp$ versus $\np$.
KN do not include these additional effects, the only input
parameters being the fraction of particle production from hard
processes and the earlier PHOBOS result.  
This leads to a dependence of $\dndeta$ on $\np$ similar to that
measured by WA98.

The second class of calculations, based on parton saturation, 
predict a different dependence on the nuclear geometry.
For example, 
the EKRT model \cite{ekrt}, which incorporates a geometry-dependent saturation
scale, predicts a near-constant dependence of $\dndetaonp$ 
as a function of $\np$.
In Ref. \cite{kn}, KN also perform a calculation based on parton saturation, 
including the DGLAP evolution of the gluon structure function.
They find that $\dndetaonp$
scales as $\ln ( Q^2_s / \Lambda^2 )$, where $Q^2_s$ is the
saturation momentum scale which depends on the impact parameter. 
Perhaps fortuitously, this calculation is in near-perfect
agreement with the other KN calculation, 
suggesting that these two physics scenarios may not be distinguishable,
even in principle.

In this Letter, we present the results of 
a measurement of the charged particle multiplicity
per participating nucleon pair near mid-rapidity,
$\dndetaonenp$, as a function of $\np$.
For this measurement, we used a subset of the full PHOBOS detector, which
was partially described in Ref. \cite{dndeta}.  

To measure the charged particle multiplicity, we used
two of the three silicon detector systems implemented in PHOBOS\cite{phobos2}, 
each of which has different properties
and thus different systematic effects on the data.  
The PHOBOS spectrometer ($\spec$) used for the 2000 data 
consists of two arms, 
one with 16 ($\specn$) and another with 6 layers ($\specp$).  
The first six layers of each sub-detector subtend $-1 < \eta < 2$
and $\Delta \phi < 7^o$ around $\phi = 0$ ($\specp$) and
$\phi = 180^o$ ($\specn$).
The innermost of these layers has 1 mm$^2$ pads
while the pads get narrower and taller as the distance from
the event vertex increases.
The PHOBOS vertex detector ($\vtx$) consists of two sets ($\vtxt$/$\vtxb$) of two 
layers which are located above and below the beam ($z$) axis.
Primarily designed to measure the vertex
$z$-position ($z_{vtx}$), 
the pads have very fine segmentation along $z$, but are
larger in the $x$ (horizontal) direction.  
For events with $\zvtx=0$, the detector covers $\Delta \phi \approx
\pm 22^{o}$ around $\phi = 90^o$ and $270^o$ and $\Delta \eta = \pm .97$
around $\eta=0$.

\begin{figure}
\begin{center}
\epsfig{file=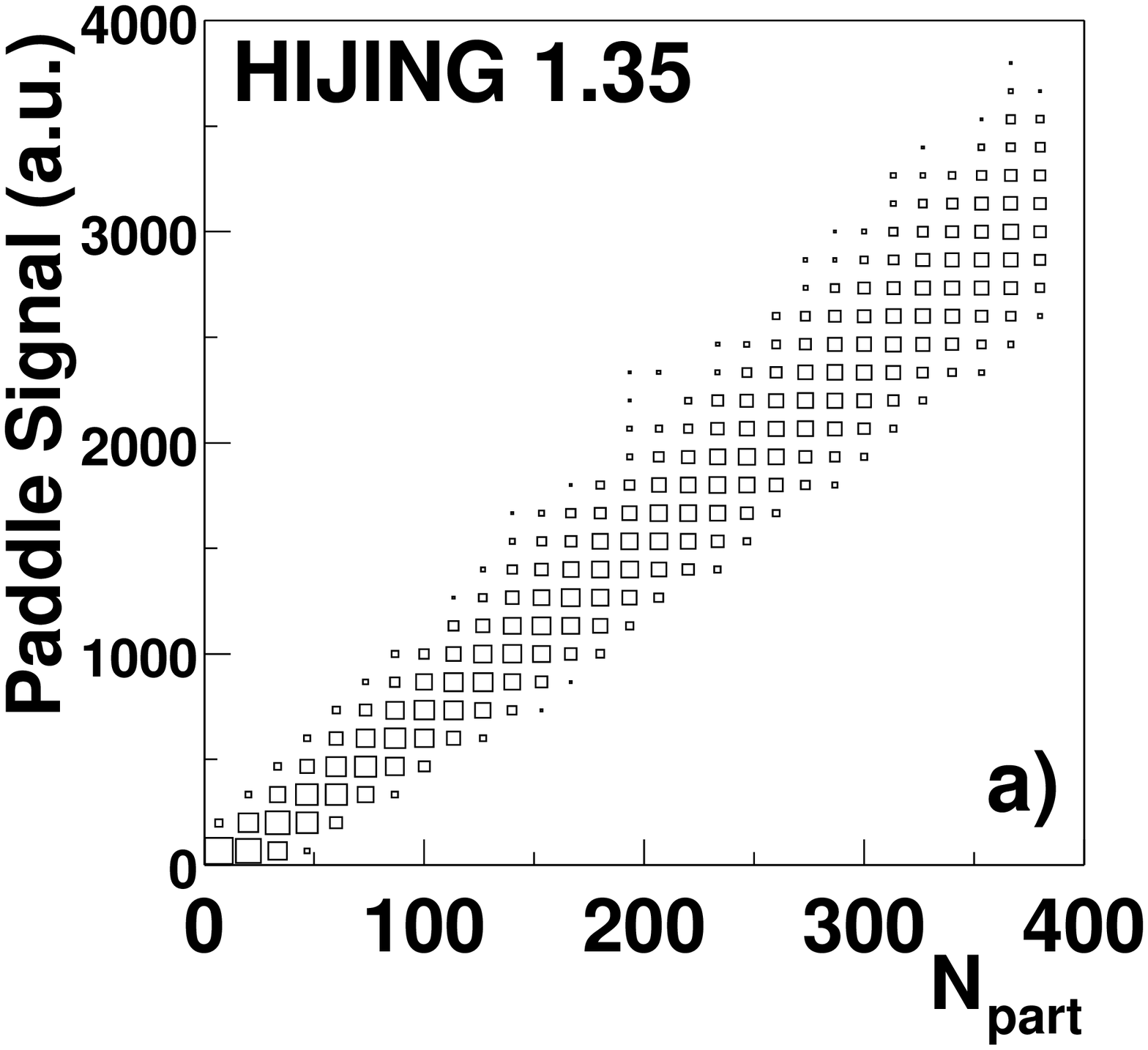,width=4cm}
\epsfig{file=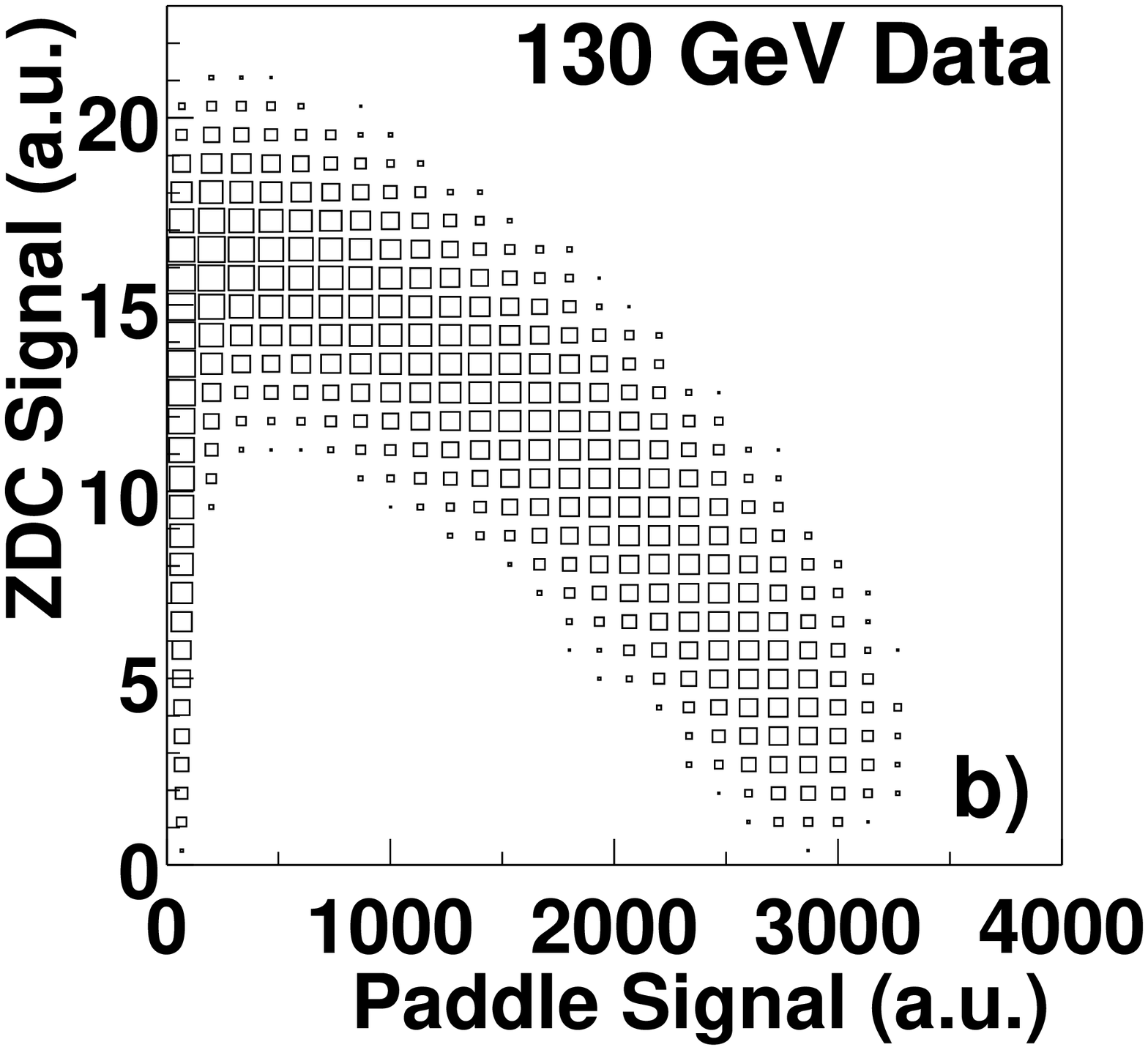,width=4cm}
\end{center}
\caption{
a.) Simulated paddle signal as a function of the number of participants.
b.) ZDC signal vs. paddle signal
for PHOBOS data at $\snn = 130$ GeV.
}
\label{pdlmean_vs_npart}
\vspace*{-.5cm}
\end{figure}

The centrality of the collisions, from which we derive $\np$,
is primarily determined using the energy
deposited by charged particles in 
two sets of paddle counters located
at $\pm 3.21$ meters from the nominal interaction point along the
beam axis, which subtend $3 < |\eta| < 4.5$.
HIJING simulations~\cite{hijing} suggest that, on average,
the paddle signal is 
monotonically related to the number of participants, as shown in
Fig. \ref{pdlmean_vs_npart}(a).
This has been verified by the PHOBOS data shown in 
Fig. \ref{pdlmean_vs_npart}(b), which shows the
correlation between the paddle signal
and the signal from the zero-degree calorimeters (ZDCs)\cite{ZDCnim}
which are located at $\pm 18$ m 
and measure the forward-going neutral spectator matter.

As a consequence of the monotonic relationship between the paddle
signal and $\np$,
fractions of the cross section as selected by the paddles
correspond on average to the same fractions of the cross section 
selected by $\np$.
To account for the fluctuations of secondaries produced in
the apparatus as well as of $\np$ itself, we actually calculate
$\avenp$ for fractions of the cross section selected using
a full simulation of the paddle response based on HIJING and
GEANT.
This has been done
for each of 10 bins in the most central 45\% of the total cross section
(shown in Table \ref{results_table}).
We find that for the PHOBOS setup, 
ignoring all sources of fluctuations leads to shifts in 
$\avenp$ of less than 2\%.

A major source of experimental systematic
error in the determination of $\avenp$ arises from
uncertainty in the efficiency 
of our event selection procedure
(described in Ref. \cite{dndeta})
for low-multiplicity events.
We have estimated this by studying the frequency distribution of
the number of hit paddle counters, which is sensitive to the most
peripheral events,
and comparing the results to Monte Carlo simulations.
By performing the
procedure with two different models (HIJING, RQMD\cite{rqmd})
we estimate a systematic error of 3\% and an efficiency of 97\%.
Unfortunately, an error of as little as 3\% 
leads to errors on $\avenp$ on the order of 5\% for $\np < 100$.  
This uncertainty 
accounts for about half of the total systematic error on the 
final result described below.

\begin{figure}[h]
\begin{center}
\epsfig{file=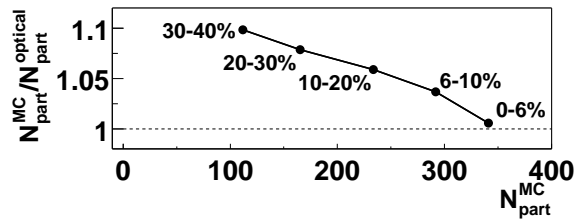,width=8cm}
\end{center}
\caption{Ratio of $\np$ calculated by KN (optical-limit approach) over $\np$ calculated by HIJING (MC approach) vs. HIJING.  Each comparison is done for the same fraction of total cross section, as indicated next to the points.
}
\label{glauber_err}
\vspace*{-.5cm}
\end{figure}

It should be noted that the Glauber model calculation 
implemented in HIJING 1.35 uses a Monte Carlo approach.
In this, nucleons are randomly distributed
according to a Woods-Saxon distribution, and interactions occur
with a probability proportional to the overlap of the Gaussian 
nucleon density profiles.
This is very similar to the procedure 
used by the PHENIX collaboration in a recent publication\cite{phenix}.
A different approach is taken by
KN \cite{kn,jpsi}, who use a
numerical integration of the nuclear overlap function that
should in principle
give identical results as the Monte Carlo approach.
However, it is well known that these calculations are done in the optical
limit to make the integrals tractable.
While this approximation is reasonable for expressing $\nch$ as 
a function of $\np$ or impact parameter, it is known to be inaccurate for
estimating the total inelastic cross section \cite{bialas}.
Thus, for the same fraction of cross section,
it can be {\it expected} to give different results 
for $\avenp$ relative to a Monte Carlo approach.
In fact, we have found that the two approaches do disagree,
and moreover, 
cannot be brought into agreement by reasonable variation of the input
parameters (e.g. radius and $\sigma_{NN}$). 
The ratio of $\np$ from HIJING (``MC'') over KN (``optical'')
is shown in Fig. \ref{glauber_err} as a function of $\np$ from
HIJING.

Following the procedure described in \cite{dndeta}, 
we have determined the charged particle multiplicity, $\dndeta$,
at midrapidity averaged over $\etaone$.
The technique is based on counting ``tracklets'', which
are three-point tracks consisting of two points and the
measured event vertex.  In this analysis, we have used combinations of
five effective ``sub-detectors'': layers 1 and 2 of $\specp$ and $\specn$,
layers 5 and 6 of $\specp$  and $\specn$, and the full vertex detector.
The large number of differently positioned
detectors allowed us to control the effects of backgrounds and
was used to check the consistency of our analysis technique.

For the spectrometer,
the pseudo-rapidity $\eta$ and azimuthal angle $\phi$
of all hits in two consecutive spectrometer layers were
calculated relative to the primary event vertex. 
Tracklets were then constructed by combining
pairs of hits in both layers for which the
total angular distance
$D = \sqrt{ \delta \eta^{2} + \delta \phi^{2} }$ 
satisfies the condition $D < 0.015$, 
where $\delta\eta$ and $\delta\phi$ are
the deviations in pseudorapidity and azimuthal angle
(in radians) of the two hits, respectively.
If two tracklets share a hit, the one with the larger value
of D is discarded.
A similar tracklet finding algorithm was used for the vertex detector.
Tracklets were chosen in the vertex detector as 
combinations of hits in the two detector layers with
$|\delta \phi| < 0.3$, and  $|\delta \eta| < 0.04$.  
We did not use the same measure $D$ as for the spectrometer since the
granularity in the vertex detector 
is substantially coarser in the $\phi$ direction.

To study the effect of combinatorial backgrounds, we analyzed
the full data sample with the inner layers of each set of detectors
rotated about the beam
axis by 180 degrees.  While preserving the gross features of the events
(e.g. flow), the tracklets extracted from this data set
arise exclusively from random coincidences of two hits that satisfy our
quality cuts.  
Outside of the cut region, we find that the distribution of track
residuals (D for the spectrometer, $\delta \eta$ for the vertex) 
for the mixed-hit tracklets closely matches 
those obtained with the true detector geometry.  By normalizing
the two distributions outside the cuts, we thus
obtain an estimate of the combinatorial background 
in the region of accepted tracklets.

The high segmentation of the spectrometer in both 
pseudorapidity and azimuthal angle
leads to a small number of background tracklets,
which is substantially larger for the vertex detector because of
its larger pads.
In both cases, the background level was found to 
scale with the number of occupied pads.
In the spectrometer, the background varied from 1\% to 15\%, depending
on occupancy.  
The final number of tracklets is corrected for this combinatorial background
with the measured distribution,
smoothed using a 2nd-order polynomial.
The coarser segmentation of the vertex detector leads to a larger contribution
from combinatorial tracklets.
However, since the scaling of the background with the number of
occupied pads is similar in data and simulation,
we use a global correction factor to take this into account.
Making an explicit correction similar to the spectrometer 
has a negligible effect (less than 1\%) on the final answer.  

The proportionality factor $\alpha$
between the number of tracklets and 
the multiplicity of primary charged particles for $\etaone$
was calculated using the results of simulations.  
Particles generated by HIJING were propagated through GEANT 3.21.
The resulting simulated signals were smeared to account for
detector resolution, and subjected to 
the same analysis chain as the real data. 

The precise manner in which these proportionality factors were
calculated and used is somewhat different in the spectrometer and
vertex detectors.
In the spectrometer, 
$\alpha$ was computed as a function of $\zvtx$ and centrality.
Since the spectrometer acceptance is forward of mid-rapidity,
we only include tracklets within a fiducial cut of
$0 < \eta <1$.
Using these proportionality factors,
the corrected tracklet multiplicity was calculated using events 
in $-4 < \zvtx < 12$ cm. 
For each centrality and vertex bin, the background fraction, which depends on
the number of occupied pads, was also averaged over the selected events
and then applied to the average number of tracklets.
In the vertex detector, 
$\alpha$ was computed for tracklets in $|\eta|<1$ as
a function of $\zvtx$ (for $|\zvtx|<12$ cm) and $N_{outer}$, 
where $N_{outer}$ is the number of hits in the outer vertex layer.
This corrects for both the reconstruction efficiency and 
the combinatorial background.
It is more difficult to distinguish these 
these two effects in the vertex detector, 
which lacks the pointing accuracy of the spectrometer.
The final estimate of $\dndetaone$ was then determined in both cases
by averaging the corrected number of measured tracklets over $\zvtx$
in each centrality bin.

The systematic error for the spectrometer is dominated by the accuracy
of the vertex determination 
and the efficiency of the tracklet reconstruction procedure 
and is estimated to be 3\%.  
The uncertainty on the combinatorial background subtraction
has been estimated to be 1\%.  Finally, the uncertainty in the
effect of non-vertex
backgrounds (which includes weak decays) has been estimated to be less
than 1\%.  
Thus, we estimate an error of 4.5\%, independent of centrality.

The vertex detector has larger systematic errors, 
due to its coarser segmentation in the $x$-direction,
which makes it less robust against contamination from non-vertex backgrounds,
such as delta electrons and weak decays.
By considering how the final result varies with changes in the
quality cuts
and event generator used, 
we estimate a final systematic error of 7.5\%.

The final value of $\dndetaone$ is based
on an average of two separate measurements, 
one combining the four spectrometer measurements
and one with the full vertex detector.
Since the systematics are very different for the
two different analysis techniques, 
we combine the two results weighted by the inverse of their
total systematic error squared to obtain the final results, which are shown
in Table \ref{results_table}.

\begin{figure}[h]
\begin{center}
\epsfig{file=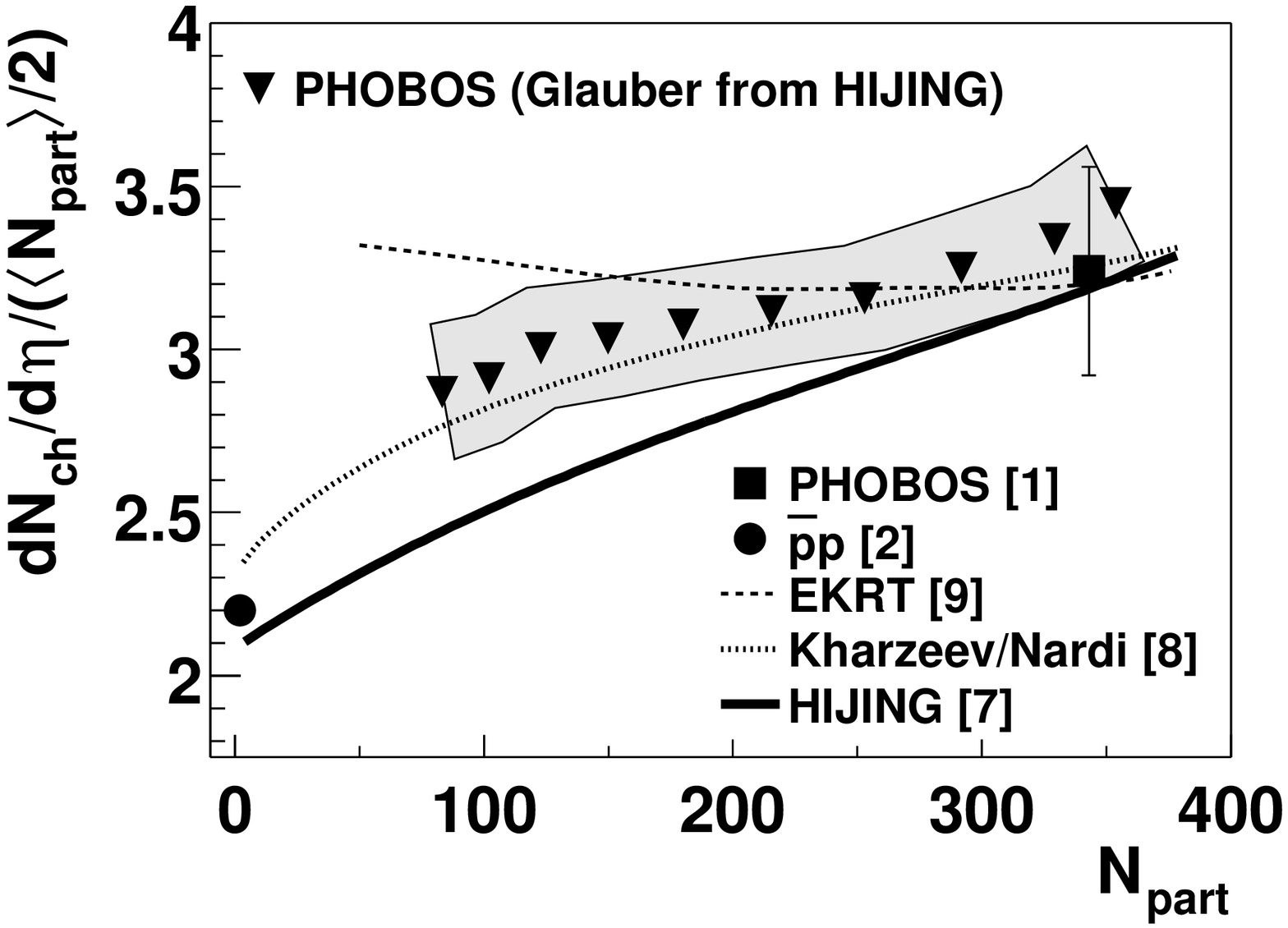,width=8cm}
\end{center}
\caption{The measured scaled pseudorapidity density $\dndetaonenp$ is shown as a function of $\np$ (solid triangles), with $\np$ extracted using HIJING.  The error band combines the error on $\dndetaone$ and $\np$.  The solid circle is $\pbarp$ data from Ref. [2].
The solid square is from Ref. [1].  Theoretical calculations are shown from HIJING~[7] (solid line), KN~[8] (dotted curve) and EKRT~[9] (dashed curve).
}
\label{final}
\vspace{-.5cm}
\end{figure}

The scaled pseudorapidity density $\dndetaone / $ $\halfnp$ 
as a function of $\np$ is shown in Fig. \ref{final},
with $\avenp$ derived using HIJING.
The two different sources of systematic error, one from the tracklet measurement
and the other from the estimation of $\np$, are combined in quadrature and shown
as a band around the data points.  The error on the participant estimation is
based on the assumption that we may have
mis-estimated the total cross section by 3\%, our
systematic error on this quantity.  
For comparison, we show an extrapolation of
$\pbarp$ measurements to $\snn=130$ GeV using
a procedure described in \cite{pp} (solid circle), as well as 
the PHOBOS measurement of the pseudorapidity density at 
midrapidity~\cite{dndeta} (solid square).
Our results are in good agreement with a recent 
PHENIX publication~\cite{phenix}.

Three model comparisons are also shown.   
The HIJING model (solid curve)
interpolates almost linearly between the 
$\overline{p}p$ point and the previous PHOBOS point\cite{dndeta}.  
Above this, we show both KN results as a single dotted curve.  
Note that the absolute scale in this model was normalized to
the PHOBOS value for the 6\% most central events \cite{kn}.
Finally, the saturation results of EKRT (dashed curve)
are nearly constant as a function of $\np$.
The data appear to disfavor the HIJING and EKRT results.
However, they broadly agree with the KN results, which are consistent
with the simplest scaling expected by a Glauber model
including a contribution proportional to $\nc$.

This work was partially supported by US DoE grants DE-AC02-98CH10886,
DE-FG02-93ER40802, DE-FC02-94ER40818, DE-FG02-94ER40865, DE-FG02-99ER41099, 
W-31-109-ENG-38 and
NSF grants 9603486, 9722606 and 0072204. 
The Polish groups were partially supported by KBN grant 2 P03B
04916. 
The NCU group was partially supported by NSC of Taiwan under 
contract NSC 89-2112-M-008-024.  We would like to thank 
A.~Bia\l as, W.~Czy\.{z}, D. Kharzeev, K. Reygers, and K. Zalewski 
for helpful discussions.

\end{multicols}
\newpage

\begin{table}
\caption{For each measured centrality bin, based on percentile of the total cross section, we show $\dndetaone$, the number of participants, and the final result for $\dndetaonenp$, including the full error estimation.  \label{results_table} } 
 \begin{tabular}{cc|cc} 
\multicolumn{2}{c}{\bf measured}&\multicolumn{2}{c}{\bf derived} \\ 
Bin(\%) & $\dndetaone$  & $\langle \np \rangle$ & $\dndetaonenp$ \\ \hline 0 - 3 & 609 $\pm$ 24 & 353 $\pm$ 11 & 3.45 $\pm$ 0.18 \\ 
 3 - 6 & 549 $\pm$ 21 & 329 $\pm$ 9 & 3.34 $\pm$ 0.16 \\ 
 6 - 10 & 474 $\pm$ 18 & 291 $\pm$ 8 & 3.25 $\pm$ 0.16 \\ 
 10 - 15 & 399 $\pm$ 15 & 252 $\pm$ 8 & 3.16 $\pm$ 0.16 \\ 
 15 - 20 & 335 $\pm$ 13 & 215 $\pm$ 7 & 3.12 $\pm$ 0.16 \\ 
 20 - 25 & 277 $\pm$ 10 & 180 $\pm$ 6 & 3.08 $\pm$ 0.17 \\ 
 25 - 30 & 227 $\pm$ 9 & 149 $\pm$ 6 & 3.03 $\pm$ 0.18 \\ 
 30 - 35 & 184 $\pm$ 7 & 122 $\pm$ 5 & 3.00 $\pm$ 0.18 \\ 
 35 - 40 & 148 $\pm$ 6 & 102 $\pm$ 5 & 2.91 $\pm$ 0.20 \\ 
 40 - 45 & 119 $\pm$ 4 & 83 $\pm$ 4 & 2.87 $\pm$ 0.21 \\ 
\end{tabular}

\end{table}

\end{document}